\documentclass[twoside,english,british,aip,apl,reprint]{revtex4-1}
\usepackage[T1]{fontenc}
\usepackage[latin9]{inputenc}
\usepackage{geometry}
\usepackage{float}
\usepackage{units}
\usepackage{amstext}
\usepackage{amssymb}
\usepackage{graphicx}

\makeatletter

\@ifundefined{textcolor}{}
{%
 \definecolor{BLACK}{gray}{0}
 \definecolor{WHITE}{gray}{1}
 \definecolor{RED}{rgb}{1,0,0}
 \definecolor{GREEN}{rgb}{0,1,0}
 \definecolor{BLUE}{rgb}{0,0,1}
 \definecolor{CYAN}{cmyk}{1,0,0,0}
 \definecolor{MAGENTA}{cmyk}{0,1,0,0}
 \definecolor{YELLOW}{cmyk}{0,0,1,0}
}

\date{31. August 2014}

\makeatother

\usepackage{babel}
\begin{document}

\title{Room Temperature Electrical Detection of Spin Polarized Currents
in Topological Insulators}

\author{André Dankert}

\email{andre.dankert@chalmers.se}

\selectlanguage{british}%

\affiliation{Department of Microtechnology and Nanoscience, Chalmers University
of Technology, Quantum Device Laboratory; Göteborg, Sweden}

\author{Johannes Geurs}

\affiliation{Department of Microtechnology and Nanoscience, Chalmers University
of Technology, Quantum Device Laboratory; Göteborg, Sweden}

\author{M. Venkata Kamalakar}

\affiliation{Department of Microtechnology and Nanoscience, Chalmers University
of Technology, Quantum Device Laboratory; Göteborg, Sweden}

\author{Saroj P. Dash}

\email{saroj.dash@chalmers.se}

\selectlanguage{british}%

\affiliation{Department of Microtechnology and Nanoscience, Chalmers University
of Technology, Quantum Device Laboratory; Göteborg, Sweden}

\maketitle
\textbf{Topological insulators (TIs) are a new class of quantum materials
that exhibit spin momentum locking (SML) of massless Dirac fermions
in the surface states.}\textbf{ Usually optical methods, such as
angle and spin-resolved photoemission spectroscopy, have been employed
to observe the helical spin polarization in the surface states of
three-dimensional (3D) TIs up to room temperatures. Recently, spin
polarized surface currents in 3D TIs were detected by electrical methods
using ferromagnetic (FM) contacts in a lateral spin-valve measurement
geometry.}\textbf{ However, probing the spin texture with such electrical
approaches is so far limited to temperatures below \mbox{125$\,$K},
which restricts its application potential.}\textbf{ Here we demonstrate
the room temperature electrical detection of the spin polarization
on the surface of Bi$_{2}$Se$_{3}$ due to SML by employing spin
sensitive FM tunnel contacts. The current-induced spin polarization
on the Bi$_{2}$Se$_{3}$ surface is probed at room temperature by
measuring a spin-valve signal while switching the magnetization direction
of the FM detector. The spin signal increases linearly with current
bias, reverses sign with current direction, exhibits a weak temperature
dependence and decreases with higher TI thickness, as predicted theoretically.
Our results demonstrate the electrical detection of the spin polarization
on the surface of 3D TIs, which could lead to innovative spin-based
quantum information technology at ambient temperatures.}

\textbf{}

The strong spin-orbit (SO) coupling in three-dimensional (3D) TIs
leads to an insulating bulk and conducting surface states protected
by time reversal symmetry \cite{Xia2009,Ando2013a,Hasan2010}. Electrons
populating these topological surface states (TSS) have their spin
and momentum locked at a right angle \cite{Hasan2010,Pesin2012a}.
These TSS have only one spin state in contrast to the two spin states
per momentum state in conventional materials. The TSS are extremely
robust against most perturbations from defects or impurities and can
enable the propagation of dissipationless spin currents \cite{Ando2013a,Hasan2010}.
Semiconducting Bi$_{2}$Se$_{3}$ is a 3D TI with a single Dirac cone
at the Fermi level \cite{Xia2009,Zhang2009a}, which makes it an ideal
prototype to study topological effects. Additionally, its weak electron-phonon
coupling allows the persistence of topological surface states up to
room temperature. Such spin helicity of the Bi$_{2}$Se$_{3}$ surface
has been experimentally measured by spin-resolved photoemission spectroscopy
up to $\unit[300]{K}$ \cite{Pan2011,Hsieh2009,Pan2012}. These optical
techniques have the advantage to probe only the surface effects without
bulk contributions. 
\begin{figure*}
\begin{centering}
\includegraphics[scale=1.2]{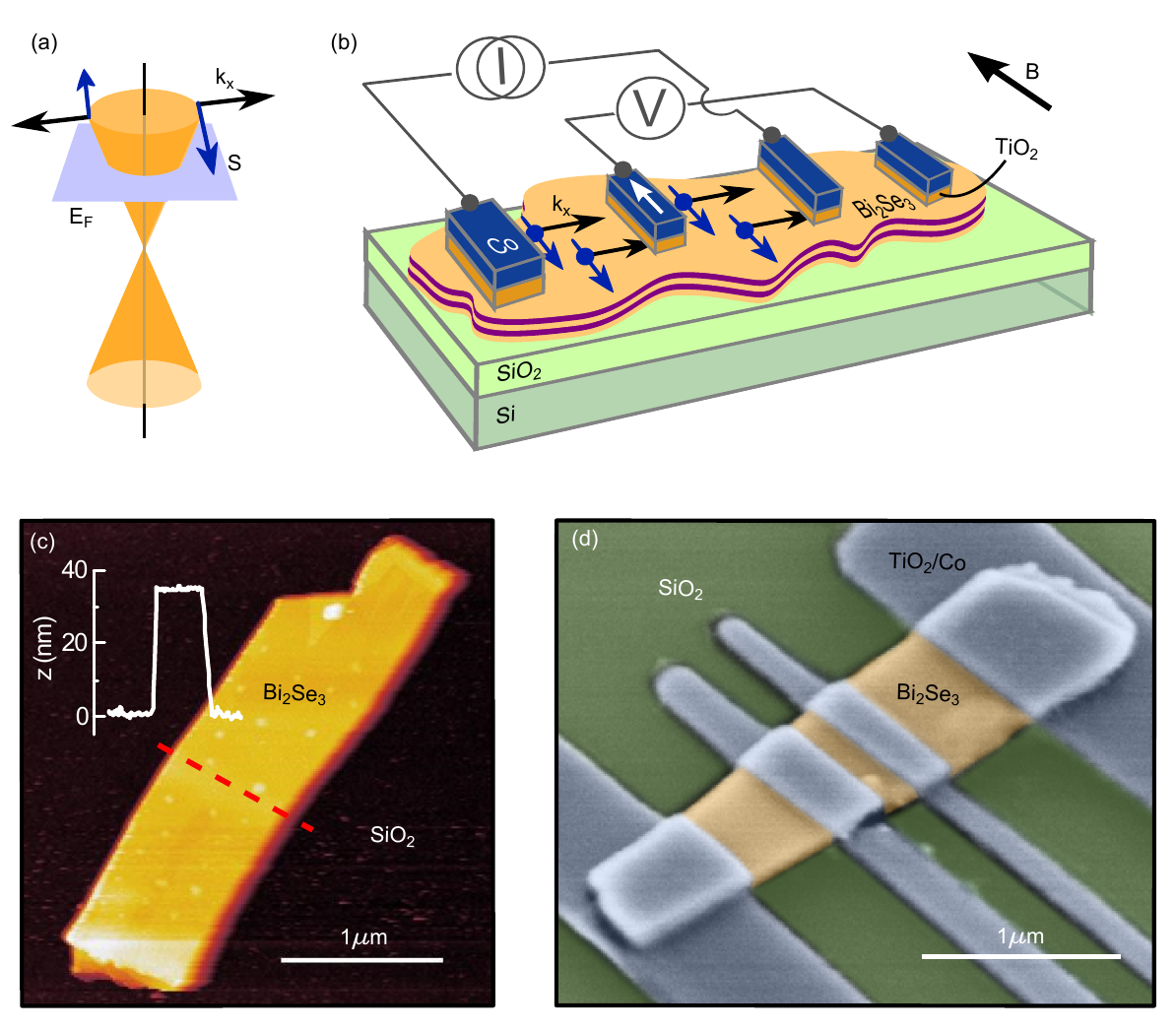} 
\par\end{centering}

\protect\caption{\textbf{\small{}Schematics of spin-valve device to probe the spin
helical topological surface states.}{\small{} (a) Dirac cone of the
topological surface state with elevated Fermi level with spins locked
perpendicular to its momentum. (b) Schematic of a TI with FM tunnel
contacts. The direction of spin current in Bi$_{2}$Se$_{3}$ is defined
by the charge current direction due to SML. (c) Atomic force microscope
image of a Bi$_{2}$Se$_{3}$ flake with a thickness of $\unit[35]{nm}$
(Inset: Line-scan (red) over the flake). (d) Coloured electron microscope
image of a Bi$_{2}$Se$_{3}$ device with FM tunnel contacts taken
after measurements.}}

\label{DevFab} 
\end{figure*}

However, the electrical detection of spin polarizations in 3D TIs
remained challenging. Undesired doping and low TI bulk band gaps usually
create a parallel unpolarized conduction channel. Therefore, the electrical
quantum spin Hall method, used for detection of spin edge states in
two-dimensional (2D) TIs \cite{Konig2007,Brune2010}, cannot be utilized
for the 3D case. So far, dynamical methods were employed to couple
the TSS to FM contacts creating a spin transfer torque \cite{Mellnik2014}
or for spin injection \cite{Shiomi2013,Deorani2014}. Only recently,
potentiometric measurements have been used to detect spin-polarized
surface currents in 3D TIs probed by a FM contact \cite{Li2014b,Tian2014,Tang2014},
which act as an efficient detector even in presence of unpolarized
bulk charge currents \cite{Burkov2010}. However, the direct electrical
detection of a current-induced spin polarization in 2D and 3D TIs
has been so far restricted to temperatures below $\unit[125]{K}$
\cite{Konig2007,Li2014b,Tian2014,Tang2014,Brune2010}, which limits
further progress in this research field and its application potentials.
The room temperature electrical detection of such highly correlated
spin systems is not only interesting for fundamental research but
also for applications in dissipationless quantum spintronic devices
\cite{Hasan2010,Pesin2012a}.

Here we demonstrate the room temperature electrical detection of spin
polarized currents on the surface of the Bi$_{2}$Se$_{3}$. An applied
electric field creates a finite momentum $\vec{k_{x}}$ of the charge
carriers yielding a perpendicularly locked spin transport $\vec{S}$
on the surface, as depicted in Fig. \ref{DevFab}a. We probe this
current-induced spin polarization by using sensitive FM tunnel contacts
(Fig. \ref{DevFab}b) resulting in a spin-valve effect between the
magnetization of the FM and the spin-polarized current on the Bi$_{2}$Se$_{3}$
surface. We used Bi$_{2}$Se$_{3}$ flakes mechanically exfoliated
from single crystals onto a SiO$_{2}$/Si substrate (Fig. \ref{DevFab}c).
The multi-terminal devices (Fig. \ref{DevFab}d) with FM tunnel contacts
of Co/TiO$_{2}$ or standard contacts of Ti/Au were prepared by electron
beam lithography (see Methods).

\begin{figure*}[t]
\begin{centering}
\includegraphics[scale=1.2]{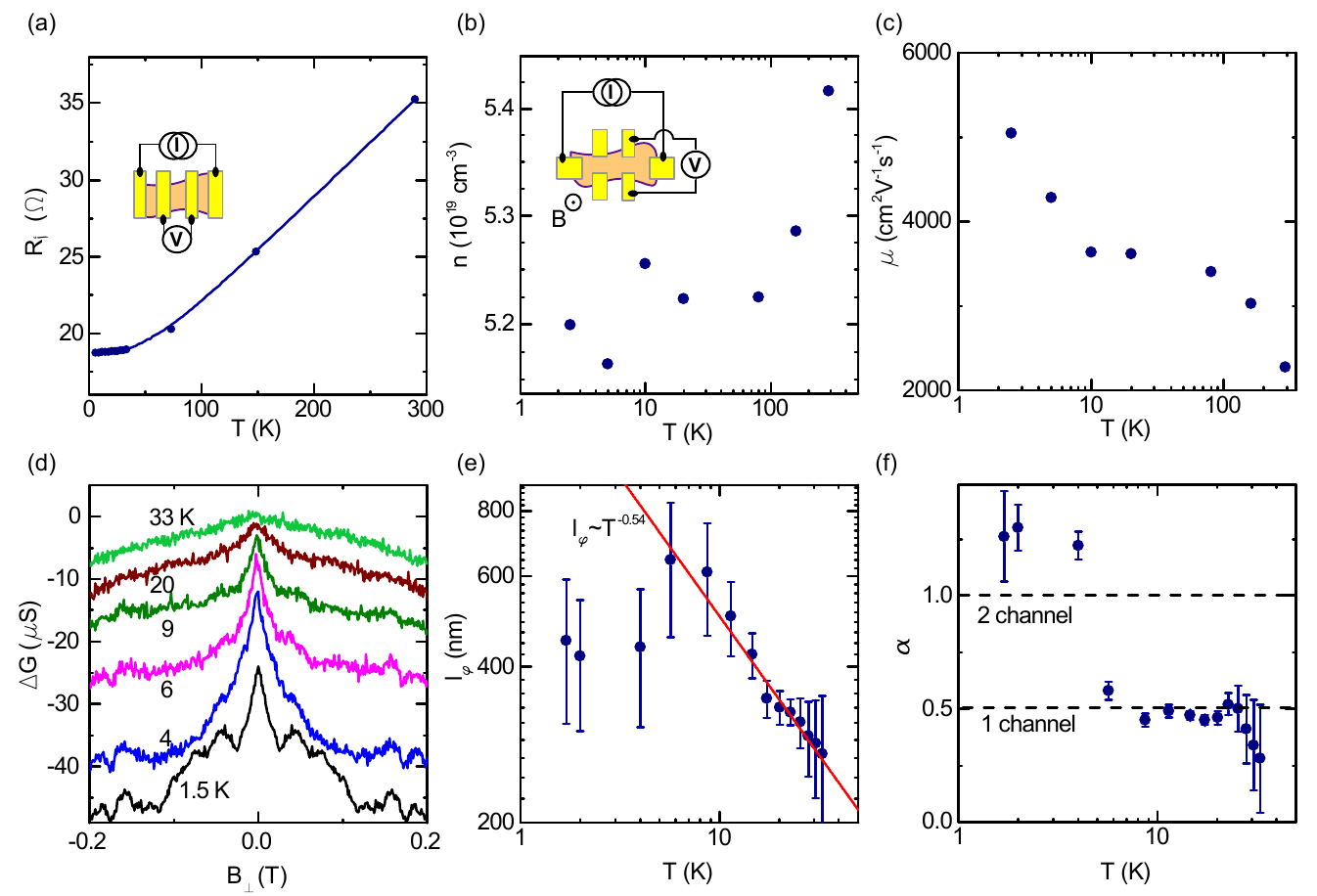} 
\par\end{centering}

\protect\caption{\textbf{\small{}Electrical and magneto-transport measurements in Bi$_{2}$Se$_{3}$.}{\small{}
(a) Temperature dependence of the Bi$_{2}$Se$_{3}$ channel resistance.
(b) }The charge carrier concentration $n\approx\unit[5\cdot10^{19}]{cm^{-3}}$
is almost constant in the studied temperature range. {\small{}(c)
Extracted Hall mobility increases strongly with decreasing temperature.
(d) Magnetoconductance measurements of the Bi$_{2}$Se$_{3}$ with
applied perpendicular magnetic field showing weak anti-localization
up to $\unit[33]{K}$. (e) Temperature dependence of channel number
$\alpha$ and (f) phase coherence length $l_{\varphi}$. }}

\label{WAL} 
\end{figure*}
The Bi$_{2}$Se$_{3}$ flakes were characterized via electrical and
magneto-transport measurements using Ti/Au contacts. The Bi$_{2}$Se$_{3}$
channel resistivity shows a metallic behaviour with a reduction in
sheet resistance $R_{\square}=\unit[36]{\Omega}$ by a factor of two
at low temperature (Fig. \ref{WAL}a). This stems from a large metallic
charge carrier concentration of $\unit[5\cdot10^{19}]{cm^{-3}}$ as
extracted from Hall measurements (Fig. \ref{WAL}b). We observe a
high channel mobility of $\unit[2000]{cm^{2}(Vs)^{-1}}$ at room temperature,
which almost doubles at $\unit[2]{K}$ (Fig. \ref{WAL}c). These results
imply a degenerate semiconducting behaviour with parallel surface
and bulk conduction channels \cite{Butch2010}. Nevertheless, it has
been demonstrated that the SML can still be probed despite a large
bulk contribution\cite{Li2014b}. Furthermore, the high mobility in
our exfoliated flakes is promising for the detection of SML \cite{Butch2010}.
These excellent properties are also reflected in the magneto-transport
measurements, where we observed weak anti-localization (WAL) up to
$\unit[33]{K}$ (Fig. \ref{WAL}d and supplementary Fig. S1) \cite{Hikami1980}.
Topological insulators exhibit a very strong SO coupling and belong
to the so-called symplectic class, where prefactor $\alpha$ should
be $\frac{1}{2}$ for one topological surface and $1$ for a bottom
and top surface\cite{He2011}. As shown in Fig. \ref{WAL}e, $\alpha$
is found to be about 1.25 below $\unit[\text{4}]{K}$ and about $0.5$
at higher temperatures. This transition coincides with a phase coherence
length $l_{\varphi}\approx\unit[400]{nm}$ at low temperatures and
a $l_{\varphi}\propto T^{-0.54}$ dependence for higher temperatures
(Fig. \ref{WAL}f). Both observations indicate that our samples are
2D systems with surface transport channels and large SO coupling \cite{He2011,Altshuler1982}.
 Such a large signature for WAL has been predicted to allow the direct
detection of surface states by transport measurements at room temperature\cite{Hsieh2009}.

We focused on the detection of SML in the surface states of these
Bi$_{2}$Se$_{3}$ flakes by FM tunnel contacts of Co/TiO$_{2}$.
The thin TiO$_{2}$ layer at the interface introduces a contact resistance
of $\unit[R_{C}\approx3]{k\Omega\mu m^{2}}$ preventing leakage current
into the FM and protecting the SML by separating the Co and the Bi$_{2}$Se$_{3}$
surface states. Such contacts are known to be very sensitive to spin
polarized currents up to room temperature, despite a background charge
current \cite{Fert2008,Dash2009}. 
\begin{figure*}[!t]
\begin{centering}
\includegraphics[scale=1.2]{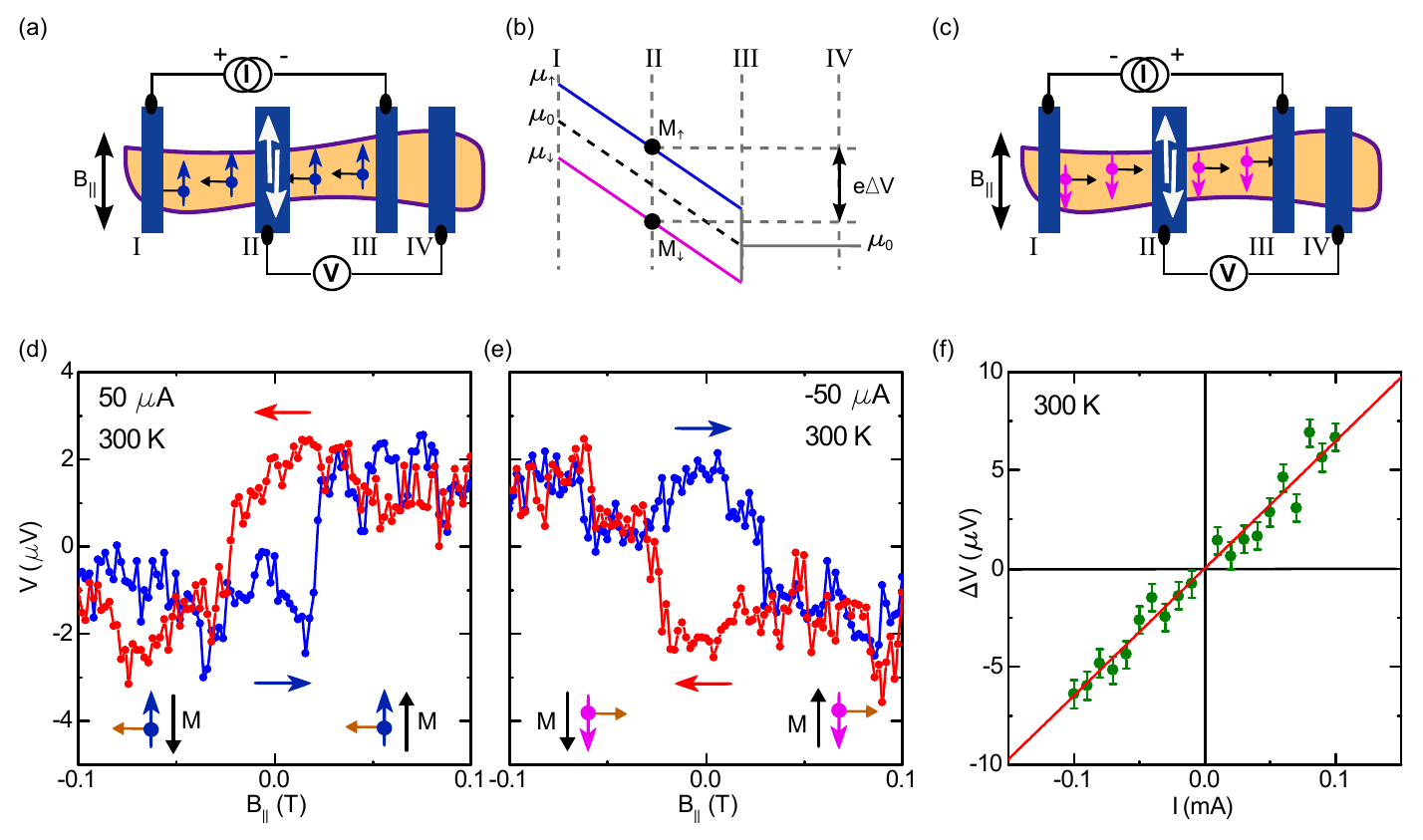} 
\par\end{centering}

\protect\caption{\textbf{\small{}Electrical detection of SML in Bi$_{2}$Se$_{3}$
at room temperature.}{\small{} (a) Schematics for spin-valve measurements
between spins ($S$) on }the Bi$_{2}$Se$_{3}$ surface{\small{} due
to SML and magnetization ($M$) of the detector FM contact. A positive
charge current is applied between contact I and III to create $S_{\uparrow}$
and a voltage signal is detected with a FM contact (II) against a
reference contact (IV). (b) The FM contact (II) measures the difference
in spin chemical potential of the channel ($\Delta\mu=\mu_{\uparrow}-\mu_{\downarrow}=e\Delta V$)
for parallel ($M_{\uparrow}$) and anti-parallel ($M_{\downarrow}$)
alignment of the FM. (c) Reversing the current direction yields a
change in spin orientation $S_{\downarrow}$.  (d) Spin-valve measurement
of Dev1 with a hysteretic switching at $\unit[300]{K}$ for a bias
current of $\unit[+50]{\mu A}$ and (e) $\unit[-50]{\mu A}$. The
arrows show the field sweep directions up (blue) and down (red). The
FM magnetization $M$ is defined by the in-plane magnetic field and
the spin $S$ defined by the current direction. (f) Bias current dependence
of the spin signal amplitude $\Delta V=V_{\uparrow\uparrow}-V_{\uparrow\downarrow}$
measured at $\unit[300]{K}$ with a linear fit (red line).}}

\label{Principle} 
\end{figure*}
Figure \ref{Principle}a shows the measurement principle of our multi-terminal
devices.  The application of an electric bias between two contacts
(I and III) results in a charge carrier flow with their spins locked
perpendiculars to their momentum. This results in an spin polarized
current on the surfaces of Bi$_{2}$Se$_{3}$. Another FM electrode
(II) is used to probe the spin potentials in Bi$_{2}$Se$_{3}$ with
respect to a reference contact (IV) outside the applied electric potential.
If the spins on the Bi$_{2}$Se$_{3}$ and the FM magnetization are
parallel, a larger potential drop can be detected than for the anti-parallel
configuration (Fig. \ref{Principle}b). The magnetization of the FM
detector can be switched by an in-plane magnetic field, whereas the
spin orientation in Bi$_{2}$Se$_{3}$ can be flipped by inverting
the current direction (Fig. \ref{Principle}c). Figure \ref{Principle}d
shows the magnetic field dependence of the detected voltage signal
at room temperature with a Bi$_{2}$Se$_{3}$ flake thickness of $\unit[35]{nm}$
(Dev1). At a fixed applied bias current of $I=\unit[+50]{\mu A}$
we observe a change in voltage signal with the change in magnetization
direction of the FM detector electrode. By sweeping the in-plane magnetic
field we obtain a hysteretic spin signal which is an analogue of a
spin-valve measurement. The switching field corresponds to the coercive
field of our Co electrode as verified by anisotropic magnetoresistance
measurements (see supplementary Fig. S3). The observed spin-valve
signal is due to the projection of the spin polarized surface current,
controlled by the current direction, on the magnetization of the FM
detector, controlled by the external in-plane magnetic field. Reversing
the current direction ($\unit[-50]{\mu A}$) locks the spins in the
opposite direction resulting in an inverted hysteretic behaviour of
the measured signal (Fig. 3e). A linear background was subtracted
from the data due to the contribution from the charge electrochemical
potential. A similar spin-valve behaviour has been measured at different
bias currents at room temperature. We observe a point symmetry around
zero bias and linear dependence of the spin signal $\Delta V$ (Fig.
\ref{Principle}d), since the spin polarization is expected to be
current independent and thus the spin density scales linearly with
the current density \cite{Hong2012}. 

From the bias dependence and different control experiments at room
temperature we can rule out any artefacts in the spin signals. The
observed change in resistance implies that the signal originates from
the SML in the TSS and not from Rashba SO coupling, which would exhibit
the opposite sign \cite{Yazyev2010a}. Any contribution of spin injections
from other FM electrodes should be suppressed by the SML in the surface
and the large SO coupling in the bulk states or would result in multiple
switchings (suppelmentary Fig. S4). The linear bias dependence also
rules out any heating related effects by the applied bias current,
which should also be independent of the sign \cite{Dankert2013a}.
Furthermore, any localized Hall effects produced by fringe fields
on the FM edges would be independent of the changes in magnetization
and current direction \cite{Li2014b}. Finally, we also rule out any
similar contribution from Lorentz magnetoresistance of the Bi$_{2}$Se$_{3}$
channel and anisotropic ferromagnetic effect of the FMs by using control
devices with $\unit[5]{nm}$ non-magnetic Ti layer at the interface
together with Co electrodes (Supplementary information S2). These
control experiments unambiguously prove that the observed signals
are due to the spin polarization in TI, which is detected by the FM
tunnel contacts.

From the SML characteristics obtained in our devices, we can analytically
evaluate the results to extract the spin polarization on the Bi$_{2}$Se$_{3}$
surface \cite{Hong2012}. The spin-valve signal $\Delta V$, induced
by a current $I_{S}$, is directly related to the surface spin polarization
$P_{S}$: 
\[
\Delta V=I_{S}R_{B}P_{S}P_{FM},
\]
 where $P_{FM}$ is the polarization of the FM detector and $R_{B}$
is the ballistic resistance \cite{Hong2012}. The ballistic conductance
$\frac{1}{R_{B}}$ can be calculated by the conductance per channel
$\frac{q^{2}}{h}$ times the number of channels $\frac{k_{\text{F}}W}{\pi}$,
where $W$ is the width of our flake. The Fermi wave number $k_{\text{F}}$
can be derived from the charge carrier concentration, which we extracted
from the Hall measurements \cite{Kittel2005}: $k_{\text{F}}=\sqrt[3]{3\pi^{2}n}.$
Furthermore, we assume a maximum FM polarization $P_{FM}\approx20\%$
for our Co/TiO$_{2}$ contacts\cite{Dankert2014a} and a surface channel
thickness of $\unit[2]{nm}$ \cite{He2010}. Under these assumptions,
we get a lower limit for the surface polarization $P_{S}\approx26\%$.
This is comparable to previous reports \cite{Li2014b} and lies well
within theoretical expectations, which predicted a limitation of the
spin polarization up to 50\% due to strong SO entanglement \cite{Yazyev2010a}.

\begin{figure*}[!t]
\begin{centering}
\includegraphics[scale=1.2]{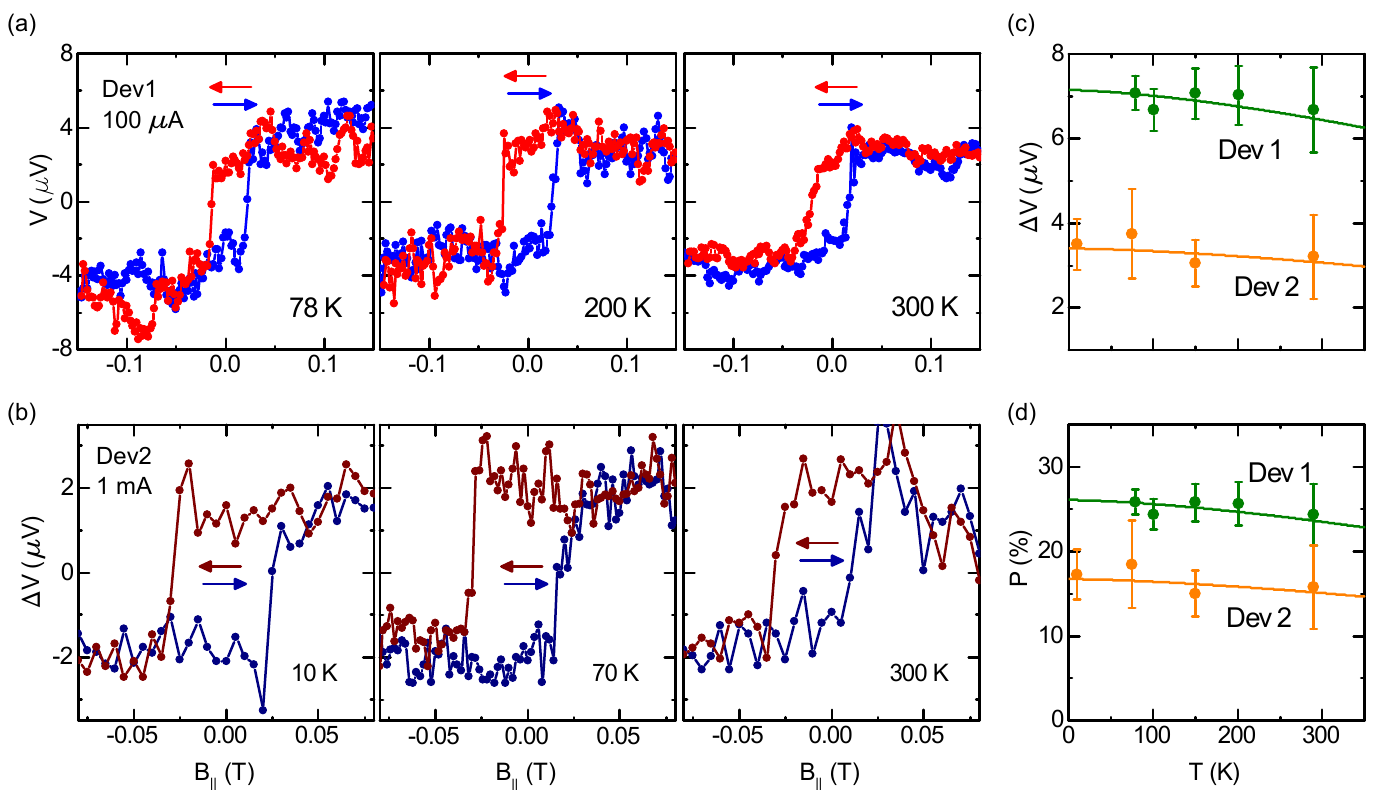} 
\par\end{centering}

\protect\caption{\textbf{\small{}Temperature and thickness dependence of the spin signal.}{\small{}
(a) Spin-valve measurements of Dev1 ($\unit[40]{nm}$ Bi$_{2}$Se$_{3}$)
at different temperatures for $\unit[100]{\mu A}$ bias current. (b)
Spin-valve measurements of Dev2 ($\unit[70]{nm}$ Bi$_{2}$Se$_{3}$)
at different temperatures for $\unit[1]{mA}$ bias current. (c) Temperature
dependence of the spin signal amplitude $\Delta V$ for two different
thicknesses (Dev1 and Dev2) of Bi$_{2}$Se$_{3}$. (d) Temperature
dependence of the spin polarization $P$ for both Bi$_{2}$Se$_{3}$
devices.}}

\label{Tdep} 
\end{figure*}
The measured spin valve signal is found to be very robust with temperature.
Figure \ref{Tdep}a shows the spin valve signals for Dev1 ($\unit[40]{nm}$
Bi$_{2}$Se$_{3}$) measured at different temperatures with a bias
current of $\unit[100]{\mu A}$. The signals exhibit a clear switching
and a signal amplitude of about $\unit[7]{\mu V}$. ``Dev2'' with
$\unit[70]{nm}$ Bi$_{2}$Se$_{3}$ shows a similar spin valve signal
with an amplitude of about $\unit[4]{\mu V}$ persistent up to room
temperature, measured at a bias current of $\unit[1]{mA}$ (Fig. \ref{Tdep}b).
The amplitude of both signals is almost independent of the temperature
(Fig. \ref{Tdep}c) and can be fitted with the theoretical model,
$\Delta V\propto1-\alpha\left(T/T_{\text{C}}\right)^{\frac{3}{2}}$,
where $T_{\text{C}}$ is the Curie temperature of Co \cite{Shang1998,Dankert2013}.
This implies that the signal amplitude depends only on the magnetization
and thus spin polarization of the TiO$_{2}$/Co tunnel contact, whereas
no significant contribution stems from the Bi$_{2}$Se$_{3}$ surface
polarization. Such a behaviour has been expected, since the surface
states are thermally stable up to room temperature as shown by optical
measurement techniques \cite{Hsieh2009,Hasan2010}. Assuming the bulk
contribution increases proportionally with the number of quintuple
layers $N$ \cite{Li2014b} and taking the different flake geometries
of both devices into account, we observe a scaling of the spin-resistance-area
product $R_{S}A=\frac{\Delta V}{I}A\propto1/N$ for both of our devices.
This is also reflected in the calculation of the spin polarization
ranging from 17-26\% (Fig. \ref{Tdep}d). Such small deviations can
also be due to differences in tunnel spin polarization of the FM tunnel
contacts prepared in different runs\cite{Dankert2014a}. These results
underline the reproducibility and implies that the signal originates
from a spin polarization on the TI surface.

In summary, we have demonstrated the room temperature electrical detection
of SML on the surface of thin exfoliated Bi$_{2}$Se$_{3}$ flakes
by ferromagnetic tunnel contacts.A hysteretic spin valve signal could
be observed reproducibly up to room temperature, with a lower limit
of the surface spin polarization of 26\%. It is expected that further
control over the spin polarization can be achieved by accessing the
dominant surface transport regime of the 3D TIs, for example by increasing
the surface-to-volume ratio, compensation doping, or electric gating
\cite{Hsieh2009,Hasan2010}. Positioning the Fermi level in the bandgap
and controlling the surface carrier density will allow the electrical
tuning of the spin polarization of the surface states \cite{Burkov2010}.
Our results will pave the way for using TIs as spin polarized sources
for spintronic devices at ambient temperatures. The possibility of
coupling TIs to other materials for spin injection \cite{Modak2012,Kamalakar2014}
opens up novel avenues in spintronic device design for energy efficient
spin-logic applications in the future.

\section*{Methods}

\textbf{Fabrication }The Bi$_{2}$Se$_{3}$ flakes were exfoliated
from a bulk single crystal (from Cradley Crystals), using the conventional
cleavage technique, onto a clean SiO$_{2}$ ($\unit[285]{nm}$)/highly
doped n-type Si substrate. The flakes were identified using a combination
of optical and atomic-force microscopy (main Fig. 1c). We used multilayer
Bi$_{2}$Se$_{3}$ with a thickness in the range of $\unit[15-100]{nm}$
and widths of $\unit[1-5]{\mu m}$.  Electrodes were patterned by
electron beam lithography followed by contact deposition in an ultra-high
vacuum electron beam evaporator. Electrodes with widths $\unit[0.3-1]{\mu m}$
and channel length of $\unit[0.2-1]{\mu m}$ are used. As contact
material we used Ti/Au for the devices for Hall and WAL measurements,
and TiO$_{2}$/Co for the detection of spin-momentum locking. \foreignlanguage{english}{The
$\unit[\approx1.5]{nm}$ TiO$_{2}$ tunnel barrier for the latter
was deposited by sputtering technique, using a DC Argon/Oxygen plasma
with a Ti target. } 

\textbf{Measurement }The devices were measured with a Keithley 2400
Sourcemeter using direct current (DC). ``Dev1'' was measured in
a liquid nitrogen cryostat between $\unit[78]{K}$ and $\unit[300]{K}$.
``Dev2'' was measured in a liquid $^{4}$He cryostat with a temperature
range of $\unit[1.5-300]{K}$.

$\vphantom{1cm}$

\textbf{Acknowledgement$\quad$}The authors acknowledge the support
of colleagues at the Quantum Device Physics Laboratory and Nanofabrication
Laboratory at Chalmers University of Technology and thank Prof. Laurens
W. Molenkamp for insightful discussions. This research is financially
supported by the Nano Area of the Advance program at Chalmers University
of Technology, EU FP7 Marie Curie Career Integration grant, the Swedish
Research Council (VR) Young Researchers Grant and an EMM Nano Consortium
scholarship.

\textbf{Competing interests$\quad$}The authors declare no competing
financial interests.


\begin{thebibliography}{10}

\bibitem{Xia2009}
Xia, Y. \textit{et~al.}
\newblock {Observation of a large-gap topological-insulator class with a single
  Dirac cone on the surface}.
\newblock \textit{Nat. Phys.} \textbf{5}, 398--402 (2009).

\bibitem{Ando2013a}
Ando, Y.
\newblock {Topological Insulator Materials}.
\newblock \textit{J. Phys. Soc. Japan} \textbf{82}, 102001 (2013).

\bibitem{Hasan2010}
Hasan, M.~Z. \& Kane, C.
\newblock {Colloquium: topological insulators}.
\newblock \textit{Rev. Mod. Phys.} \textbf{82}, 3045 (2010).

\bibitem{Pesin2012a}
Pesin, D. \& MacDonald, A.~H.
\newblock {Spintronics and pseudospintronics in graphene and topological
  insulators.}
\newblock \textit{Nat. Mater.} \textbf{11}, 409--16 (2012).

\bibitem{Zhang2009a}
Zhang, H. \textit{et~al.}
\newblock {Topological insulators in Bi2Se3, Bi2Te3 and Sb2Te3 with a single
  Dirac cone on the surface}.
\newblock \textit{Nat. Phys.} \textbf{5}, 438--442 (2009).

\bibitem{Pan2011}
Pan, Z.-H. \textit{et~al.}
\newblock {Electronic Structure of the Topological Insulator Bi$_2$Se$_3$ Using
  Angle-Resolved Photoemission Spectroscopy: Evidence for a Nearly Full Surface
  Spin Polarization}.
\newblock \textit{Phys. Rev. Lett.} \textbf{106}, 257004 (2011).

\bibitem{Hsieh2009}
Hsieh, D. \textit{et~al.}
\newblock {A tunable topological insulator in the spin helical Dirac transport
  regime.}
\newblock \textit{Nature} \textbf{460}, 1101--5 (2009).

\bibitem{Pan2012}
Pan, Z.-H. \textit{et~al.}
\newblock {Measurement of an Exceptionally Weak Electron-Phonon Coupling on the
  Surface of the Topological Insulator Bi$_{2}$Se$_{3}$ Using Angle-Resolved
  Photoemission Spectroscopy}.
\newblock \textit{Phys. Rev. Lett.} \textbf{108}, 187001 (2012).

\bibitem{Konig2007}
K\"{o}nig, M. \textit{et~al.}
\newblock {Quantum spin hall insulator state in HgTe quantum wells.}
\newblock \textit{Science} \textbf{318}, 766--70 (2007).

\bibitem{Brune2010}
Br\"{u}ne, C. \textit{et~al.}
\newblock {Evidence for the ballistic intrinsic spin Hall effect in HgTe
  nanostructures}.
\newblock \textit{Nat. Phys.} \textbf{6}, 448--454 (2010).

\bibitem{Mellnik2014}
Mellnik, A.~R. \textit{et~al.}
\newblock {Spin-transfer torque generated by a topological insulator}.
\newblock \textit{Nature} \textbf{511}, 449--451 (2014).

\bibitem{Shiomi2013}
Shiomi, Y. \textit{et~al.}
\newblock {Bulk topological insulators as inborn spintronics detectors}.
\newblock \textit{arXiv:1312.7091}  (2013).

\bibitem{Deorani2014}
Deorani, P. \textit{et~al.}
\newblock {Observation of inverse spin Hall effect in bismuth selenide}.
\newblock \textit{arXiv:1404.1146}  (2014).

\bibitem{Li2014b}
Li, C. \textit{et~al.}
\newblock {Electrical detection of charge-current-induced spin polarization due
  to spin-momentum locking in Bi$_{2}$Se$_{3}$}.
\newblock \textit{Nat. Nanotechnol.} \textbf{9}, 218--224 (2014).

\bibitem{Tian2014}
Tian, J. \textit{et~al.}
\newblock {Topological insulator based spin valve devices: Evidence for spin
  polarized transport of spin-momentum-locked topological surface states}.
\newblock \textit{Solid State Commun.} \textbf{191}, 1--5 (2014).

\bibitem{Tang2014}
Tang, J. \textit{et~al.}
\newblock {Electrical Detection of Spin-Polarized Surface States Conduction in
  (Bi$_0.53$Sb$_0.47$)$_2$Te$_3$ Topological Insulator.}
\newblock \textit{Nano Lett.} 1--18 (2014).

\bibitem{Burkov2010}
Burkov, A.~A. \& Hawthorn, D.~G.
\newblock {Spin and Charge Transport on the Surface of a Topological
  Insulator}.
\newblock \textit{Phys. Rev. Lett.} \textbf{105}, 066802 (2010).

\bibitem{Butch2010}
Butch, N.~P. \textit{et~al.}
\newblock {Strong surface scattering in ultrahigh-mobility Bi$_{2}$Se$_{3}$
  topological insulator crystals}.
\newblock \textit{Phys. Rev. B} \textbf{81}, 241301 (2010).

\bibitem{Hikami1980}
Hikami, S., Larkin, A. \& Nagaoka, Y.
\newblock {Spin-orbit interaction and magnetoresistance in the two dimensional
  random system}.
\newblock \textit{Prog. Theor. Phys.} \textbf{63}, 707--710 (1980).

\bibitem{He2011}
He, H.-T. \textit{et~al.}
\newblock {Impurity Effect on Weak Antilocalization in the Topological
  Insulator Bi$_{2}$Te$_{3}$}.
\newblock \textit{Phys. Rev. Lett.} \textbf{106}, 166805 (2011).

\bibitem{Altshuler1982}
Altshuler, B.~L., Aronov, A.~G. \& Khmelnitsky, D.~E.
\newblock {Effects of electron-electron collisions with small energy transfers
  on quantum localisation}.
\newblock \textit{J. Phys. C Solid State Phys.} \textbf{15}, 7367--7386 (1982).

\bibitem{Fert2008}
Fert, A.
\newblock {Nobel Lecture: Origin, development, and future of spintronics}.
\newblock \textit{Rev. Mod. Phys.} \textbf{80}, 1517--1530 (2008).

\bibitem{Dash2009}
Dash, S.~P., Sharma, S., Patel, R.~S., de~Jong, M.~P. \& Jansen, R.
\newblock {Electrical Creation of Spin Polarization in Silicon at Room
  Temperature}.
\newblock \textit{Nature} \textbf{462}, 491--494 (2009).

\bibitem{Hong2012}
Hong, S., Diep, V., Chen, Y.~P. \& Datta, S.
\newblock {Modeling potentiometric measurements in topological insulators
  including parallel channels}.
\newblock \textit{Phys. Rev. B} \textbf{86}, 085131 (2012).

\bibitem{Yazyev2010a}
Yazyev, O.~V., Moore, J.~E. \& Louie, S.~G.
\newblock {Spin Polarization and Transport of Surface States in the Topological
  Insulators Bi$_{2}$Se$_{3}$ and Bi$_{2}$Te$_{3}$ from First Principles}.
\newblock \textit{Phys. Rev. Lett.} \textbf{105}, 266806 (2010).

\bibitem{Dankert2013a}
Dankert, A. \& Dash, S.~P.
\newblock {Thermal Creation of Electron Spin Polarization in n-Type Silicon}.
\newblock \textit{Appl. Phys. Lett.} \textbf{103}, 242405 (2013).

\bibitem{Kittel2005}
Kittel, C. \& McEuen, P.
\newblock \textit{{Introduction to Solid State Physics}}, volume~10.
\newblock John Wiley and Sons, 8 edition (2005).

\bibitem{Dankert2014a}
Dankert, A., Mutta, V.~K., Bergsten, J. \& Dash, S.~P.
\newblock {Spin transport and precession in graphene measured by nonlocal and
  three-terminal methods}.
\newblock \textit{Appl. Phys. Lett.} \textbf{104}, 192403 (2014).

\bibitem{He2010}
He, K. \textit{et~al.}
\newblock {Crossover of the three-dimensional topological insulator Bi2Se3 to
  the two-dimensional limit}.
\newblock \textit{Nat. Phys.} \textbf{6}, 584--588 (2010).

\bibitem{Shang1998}
Shang, C., Nowak, J., Jansen, R. \& Moodera, J.~S.
\newblock {Temperature dependence of magnetoresistance and surface
  magnetization in ferromagnetic tunnel junctions}.
\newblock \textit{Phys. Rev. B} \textbf{58}, R2917--R2920 (1998).

\bibitem{Dankert2013}
Dankert, A., Dulal, R.~S. \& Dash, S.~P.
\newblock {Efficient Spin Injection into Silicon and the Role of the Schottky
  Barrier}.
\newblock \textit{Sci. Rep.} \textbf{3}, 3196 (2013).

\bibitem{Modak2012}
Modak, S., Sengupta, K. \& Sen, D.
\newblock {Spin injection into a metal from a topological insulator}.
\newblock \textit{Phys. Rev. B} \textbf{86}, 205114 (2012).

\bibitem{Kamalakar2014}
Kamalakar, M.~V., Dankert, A., Bergsten, J., Ive, T. \& Dash, S.~P.
\newblock {Enhanced Tunnel Spin Injection into Graphene using Chemical Vapor
  Deposited Hexagonal Boron Nitride}.
\newblock \textit{Sci. Rep.} \textbf{4}, 6146 (2014).

\bibitem{Patel2009}
Patel, R.~S., Dash, S.~P., de~Jong, M.~P. \& Jansen, R.
\newblock {Magnetic tunnel contacts to silicon with low-work-function ytterbium
  nanolayers}.
\newblock \textit{J. Appl. Phys.} \textbf{106}, 016107 (2009).

\bibitem{Jansen2010a}
Jansen, R., Min, B.-C. \& Dash, S.~P.
\newblock {Oscillatory spin-polarized tunnelling from silicon quantum wells
  controlled by electric field.}
\newblock \textit{Nat. Mater.} \textbf{9}, 133--8 (2010).

\bibitem{Tombros2007}
Tombros, N., Jozsa, C., Popinciuc, M., Jonkman, H.~T. \& van Wees, B.~J.
\newblock {Electronic Spin Transport and Spin Precession in Single Graphene
  Layers at Room Temperature}.
\newblock \textit{Nature} \textbf{448}, 571--574 (2007).

\bibitem{LeBreton2011a}
{Le Breton}, J.-C., Sharma, S., Saito, H., Yuasa, S. \& Jansen, R.
\newblock {Thermal spin current from a ferromagnet to silicon by Seebeck spin
  tunnelling.}
\newblock \textit{Nature} \textbf{475}, 82--5 (2011).

\end{thebibliography}

\newpage

\onecolumngrid
\setcounter{figure}{0}
\setcounter{section}{0}
\makeatletter  
\renewcommand{\thesection}{S\@arabic\c@section}
\renewcommand{\thefigure}{S\@arabic\c@figure}
\renewcommand{\thetable}{S\@arabic\c@table}    
\setcounter{page}{1}

\part*{\textsc{\Large{}Supplementary information}\\ \hspace{5mm} \\Room
Temperature Electrical Detection of Spin Polarized Currents in Topological
Insulators}

\author{André Dankert}

\affiliation{Department of Microtechnology and Nanoscience, Chalmers University
of Technology, Quantum Device Laboratory; Göteborg, Sweden}

\author{Johannes Geurs}

\affiliation{Department of Microtechnology and Nanoscience, Chalmers University
of Technology, Quantum Device Laboratory; Göteborg, Sweden}

\author{M. Venkata Kamalakar}

\affiliation{Department of Microtechnology and Nanoscience, Chalmers University
of Technology, Quantum Device Laboratory; Göteborg, Sweden}

\author{Saroj P. Dash}

\affiliation{Department of Microtechnology and Nanoscience, Chalmers University
of Technology, Quantum Device Laboratory; Göteborg, Sweden}

\section*{}

\section{Weak anti-localization measurement in Bi$_{2}$Se$_{3}$}

\begin{figure}[H]
\begin{centering}
\includegraphics[scale=1.2]{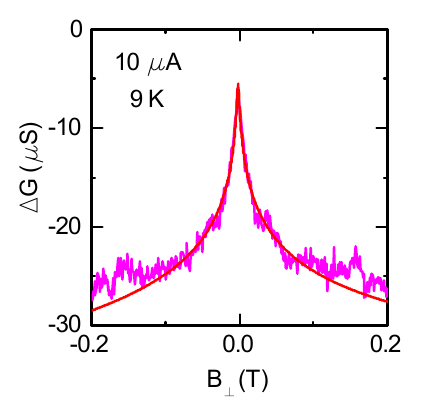} 
\par\end{centering}

\protect\caption{\textbf{\small{}Weak-antilocalization measurement and fitting.}{\small{}
}Magnetoresistance of Bi$_{2}$Se$_{3}$ channel measured at $\unit[10]{\mu A}$
showing WAL applying an perpendicular magnetic field, fitted with
Eq. (\ref{eq:WAL-1}) (red line). }

\label{WAL-1} 
\end{figure}

Weak anti-localization is an interaction of an electron wave function
in a material with a strong spin-orbit (SO) coupling. Without a SO
interaction, a time-reversal pair of electron waves, scattered by
impurities, interferes destructively (electron localization). This
quantum interference occurs at the scale of the coherence length and
yields a correction reducing the conductance around zero magnetic
field. In comparison, a strong SO interaction enhances the conductance,
which is known as the weak anti-localization (WAL) effect (Fig. \ref{WAL-1}).
In the case of a 2D system, increasing an out-of-plane magnetic field
$B_{\perp}$breaks the time-reversal symmetry, which reduces gradually
the conductance correction. This reduction can be explained by the
Hikami-Larkin-Nagaoka (HLN) model\cite{Hikami1980}. In this model,
the conductance correction $\Delta\sigma\left(B\right)$ is given
by

\begin{equation}
\Delta\sigma\left(B_{\perp}\right)=\alpha\frac{e^{2}}{\pi h}\left[\text{ln}\frac{\hbar}{4el_{\varphi}B_{\perp}}-\psi\left(\frac{1}{2}+\frac{\hbar}{4el_{\varphi}B_{\perp}}\right)\right],\label{eq:WAL-1}
\end{equation}
where $\psi$ represents the digamma function, $l_{\varphi}$ is the
phase coherence length, and the prefactor $\alpha$ describes the
quantum system\cite{Hikami1980}. The decreasing peak size with increasing
temperature is characteristic for WAL and we could observe a signal
to a, for TIs unusually high, temperature of $\unit[33]{K}$ \cite{He2011}.
Three-dimensional TIs with 2D surface states usually exhibit a very
strong SO coupling and belong to the so-called symplectic class, where
$\alpha$ should be $\frac{1}{2}$ for one topological surface and
$1$ for a bottom and top surface\cite{He2011}. Figure \ref{WAL-1}\textbf{\textsc{\Large{}
}}shows the WAL measurement at $\unit[9]{K}$ and the fitting with
the HLN model, identical to the one in the main text.

\section{Control experiments with non-magnetic contacts}

\begin{figure}[H]
\begin{centering}
\includegraphics[scale=1.2]{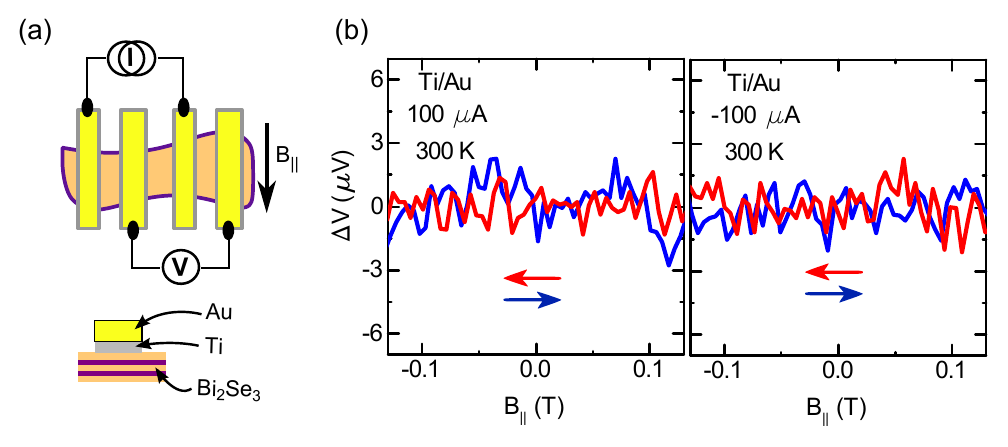} 
\par\end{centering}

\protect\caption{\textbf{\small{}Control experiment with non-magnetic contacts.}{\small{}
}(a) Multi-terminal measurement configuration using Ti/Au contacts
on a Bi$_{2}$Se$_{3}$ flake. (b) Measurement with in-plane magnetic
field sweep resulting in no hysteretic switching.}

\label{Control} 
\end{figure}

To demonstrate that the hysteretic switching, presented in the main
paper, stems from a spin-valve effect between the current-induced
spin polarization in the surface of the Bi$_{2}$Se$_{3}$ and the
magnetization of the ferromagnetic (FM) contact, we prepared devices
with non-magnetic Ti/Au electrodes (Fig. \ref{Control}a) and with
a non-magnetic Ti interlayer between Bi$_{2}$Se$_{3}$ and Co (Fig.
\ref{Control}a). This non-magnetic interlayer is known to suppress
the spin polarization of the FM detector contact, while still keeping
its magnetic properties\cite{Patel2009,Jansen2010a}. The voltages
detected by those nonmagnetic detector contacts exhibit no such field
dependent switching, which rules out any other mechanisms,such as
Lorentz magnetoresistance of the Bi$_{2}$Se$_{3}$ channel and anisotropic
magnetoresistance (AMR) of the Co contacts (Fig. \ref{Control}b)
\cite{Dash2009,Dankert2013}. 

\begin{figure}[H]
\begin{centering}
\includegraphics[scale=1.2]{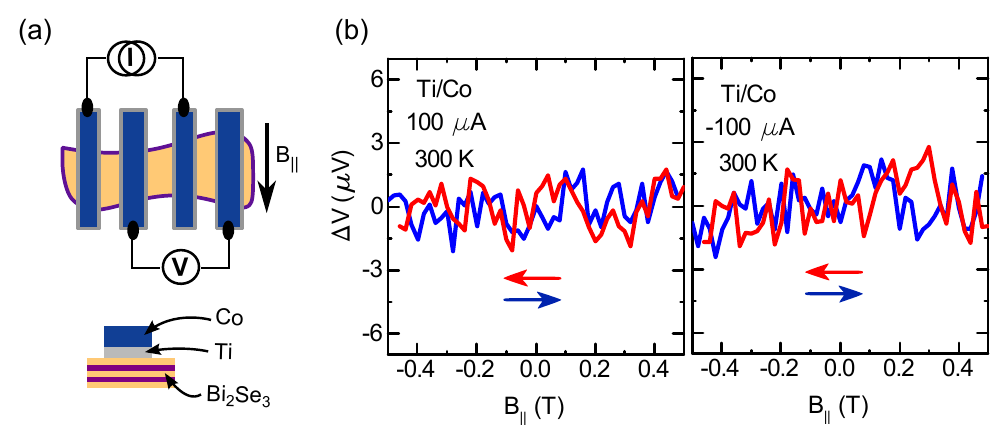} 
\par\end{centering}

\protect\caption{\textbf{\small{}Control experiment with non-magnetic interlayer.}{\small{}
}(a) Multi-terminal measurement configuration using Ti/Co contacts
on a Bi$_{2}$Se$_{3}$ flake. (b) Measurement with in-plane magnetic
field sweep resulting in no hysteretic switching.}

\label{Control2} 
\end{figure}

\section{Influence of the multi-terminal configuration}

\begin{figure}[H]
\begin{centering}
\includegraphics[scale=1.2]{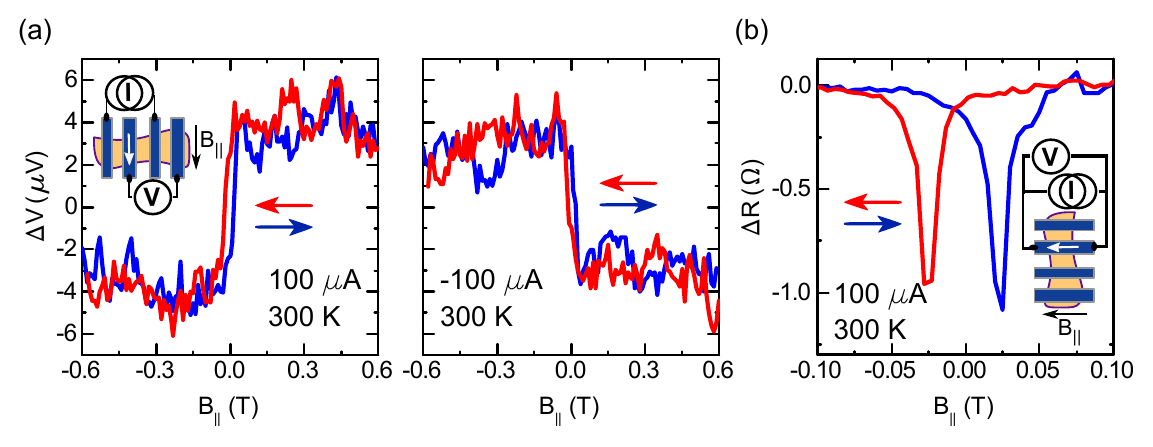} 
\par\end{centering}

\protect\caption{\textbf{\small{}High magnetic field spin-valve and anisotropic magnetoresistance
measurements.}{\small{} (a) Spin signals measured up to $\unit[\pm600]{mT}$
at $\pm\unit[100]{\mu A}$ in Dev1 at $\unit[300]{K}$ (similar to
Fig. 3 and Fig. 4 in the main text). Except for the hysteretic switching
at about $\pm\unit[25]{mT}$, no other switching events can be observed.
Inset: Measurement configuration. (b)} Anisotropic magnetoresistance
(AMR) measurement: applying a current $I$ in an identical FM contact
and measuring its resistance change while sweeping a magnetic field
parallel to the FM magnetization. The AMR signal verifies the coercive
field at $\pm\unit[25]{mT}$ for this contact, confirming the switching
field of the FM observed in the spin-valve measurement.}

\label{MultiT} 
\end{figure}

We can also rule out any contribution of spin transport in the Bi$_{2}$Se$_{3}$
channel due to spin injection from the FM electrodes in our measured
spin valves. Using a multi-terminal geometry could result in a similar
signal, if the biasing FM electrodes induce a spin accumulation in
the flake, which can be detected by the FM voltage probe. A single
switch, as observed in our case, could be the result of an incomplete
spin-valve measurement, similar to the ones in graphene \cite{Tombros2007}.
Such an effect can be ruled out, since it would result in additional
resistance changes when the injector contact magnetization is switching.
The coercive field of the FM contacts we used should be in the range
of $\unit[0-100]{mT}$. Sweeping the in-plane parallel field even
to $\unit[600]{mT}$ does only show the initially observed hysteretic
switching (Fig. S\ref{MultiT}a). Furthermore, this switching field
correlates with the coercive field of the detector contact extracted
from anisotropic magnetoresistance measurements (Fig. \ref{MultiT}b).
These observations are also supported by the theoretical expectations
of a extremely short spin lifetime in the bulk and surface states\cite{Ando2013a,Hasan2010}.
The strong spin-orbit coupling leads to a fast relaxation of any spin
accumulation limiting the diffusion length to less than the distance
between our electrodes. Therefore, only the spin polarization induced
by the spin-momentum coupling on the surface can be detected. The
origin of the spin accumulation in the surface is supported by the
thickness dependence (see ``Spin-momentum locking in thick Bi$_{2}$Se$_{3}$
(Dev2)''). Furthermore, control experiments with nonmagnetic contacts
demonstrate the spin-related nature of the signal (see ``Control
experiment'').

\section{Spin-momentum locking in thick Bi$_{2}$Se$_{3}$ (Dev2)}

\begin{figure}[H]
\begin{centering}
\includegraphics[scale=1.2]{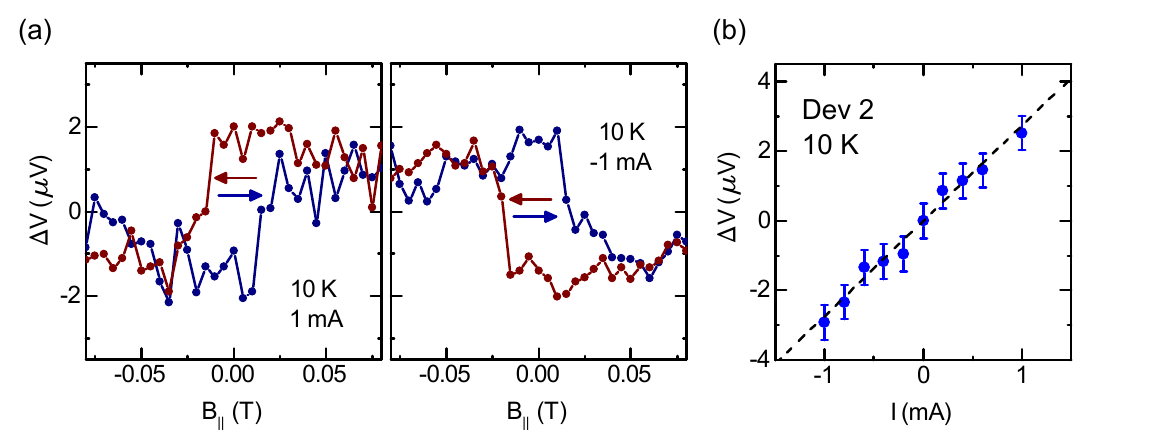} 
\par\end{centering}

\protect\caption{\textbf{\small{}Bias dependence of the SML signal of ``Dev2''.}{\small{}
The }Bi$_{2}$Se$_{3}${\small{} flake thickness is about }$\unit[70]{nm}${\small{}.
(a) Spin signal due to spin-momentum locking measured at $\unit[\pm1]{mA}$.
(b) The bias dependence of the spin signal amplitude measured at $\unit[10]{K}$
shows a linear behaviour.}}

\label{Dev2} 
\end{figure}

As mentioned in the main text, the signal was reproduced in devices
of different thicknesses. Dev1 had a thickness of about $\unit[35]{nm}$
and exhibited a clear hysteretic switching with an amplitude of about
$\unit[7]{\mu V}$ at room temperature, using $\unit[100]{\mu A}$
bias current. Dev2 had a thickness of about about $\unit[70]{nm}$
resulting in $\unit[4]{\mu V}$ signal amplitude using $\unit[1]{mA}$
bias current at room temperature (Fig. S\ref{Dev2}a). Such a behaviour
is expected, since the bulk contribution increases with increasing
number of quintuple layers $N$ resulting in a lower signal at the
same current density \cite{Li2014b}. Taking different contact areas
$A$ into account, the scaling of the spin-resistance-area product
should be $R_{S}A=\frac{\Delta V}{I}A\propto1/N$. Regarding this,
the signals measured in Dev2 scale accordingly and have a comparable
temperature (see main Fig. 4) and bias dependence to the ones of Dev1
(Fig. \ref{Dev2}b). The linear bias dependence of the spin signal
and the sign change with bias current rule out signal due to thermal
effects \cite{LeBreton2011a,Dankert2013a}.

\section*{}
\end{document}